\documentclass[twocolumn,showpacs,amsmath,amssymb,superscriptaddress,prl]{revtex4}

\usepackage{graphicx}
\usepackage{dcolumn} 
\usepackage{bm}      
\usepackage{nicefrac}
\usepackage{mathrsfs}
\usepackage[dvips]{color} 

\newcommand{\BCPO}{BiCu$_2$PO$_6$}
\newcommand{\BCPOd}{Bi(Cu$_{1-x}$Zn$_x$)$_2$PO$_6$}

\begin{document}

\title{Direct observation of impurity-induced magnetism  in an $S = \nicefrac{1}{2}$\\
antiferromagnetic Heisenberg 2-leg spin ladder}
\author{F.\ Casola}
\author{T.\ Shiroka}
\email{tshiroka@phys.ethz.ch}
\affiliation{Laboratorium f\"ur Festk\"orperphysik, ETH H\"onggerberg, CH-8093 Z\"urich, Switzerland}
\affiliation{Paul Scherrer Institut, CH-5232 Villigen PSI, Switzerland}%
\author{S.\ Wang}%
\affiliation{Laboratory for Developments and Methods, Paul Scherrer Institut, CH-5232 Villigen PSI, Switzerland}%
\affiliation{Laboratory for Quantum Magnetism, Ecole Polytechnique F\'ed\'erale de Lausanne, CH-1015 Lausanne, Switzerland}
\author{K. Conder}%
\affiliation{Laboratory for Developments and Methods, Paul Scherrer Institut, CH-5232 Villigen PSI, Switzerland}%
\author{E. Pomjakushina}%
\affiliation{Laboratory for Developments and Methods, Paul Scherrer Institut, CH-5232 Villigen PSI, Switzerland}%
\author{J.\ Mesot}%
\affiliation{Laboratorium f\"ur Festk\"orperphysik, ETH H\"onggerberg, CH-8093 Z\"urich, Switzerland}
\affiliation{Paul Scherrer Institut, CH-5232 Villigen PSI, Switzerland}%
\author{H.-R.\ Ott}
\affiliation{Laboratorium f\"ur Festk\"orperphysik, ETH H\"onggerberg, CH-8093 Z\"urich, Switzerland}

\date{\today}

\begin{abstract}
Nuclear magnetic resonance and magnetization measurements were used to probe the magnetic features of 
single-crystalline \BCPOd\ with $0<x<0.05$ at temperatures between 2.6 K and 300 K. The simple lineshape of the $^{31}$P NMR signals of the pristine compound changes considerably for $x>0$ and we present clear evidence for a 
temperature-dependent variation of the local magnetization close to the Zn sites. The generic nature of this observation is indicated by results of model calculations on appropriate spin systems of limited size employing QMC methods. 
\end{abstract}



\maketitle

Systems with coupled spin chains also termed spin ladders are currently investigated as paradigms of low-dimensional magnets, because they provide a rich playground for studies of quantum criticality and general aspects of many-body physics \cite{sachdev08}. Experiments are intended to test the predictions of model calculations or to explore new unanticipated physical features of these systems. For spin $S = \nicefrac{1}{2}$ antiferromagnetic (AFM) 2-leg ladders the ground state is a non magnetic singlet \cite{chaboussant98}. Upon application of a magnetic field the gap to the excited triplet state may be closed at $H = H_c$. A field induced Luttinger liquid (LL) phase was recently observed in this class of materials \cite{thielemann09} and theoretical predictions about its behaviour could be tested quantitatively \cite{klanjsek08}. For $H > H_c$ and at low temperatures, interladder coupling may cause the onset of magnetic order, forbidden in the range of $H < H_c$ by quantum fluctuations \cite{wessel00}. Local enhancement of magnetic correlations is also predicted to be induced by nonmagnetic entities residing on regular spin sites (spin holes) 
also for $H < H_c$ and it has been argued that studies of the influence of such defects in spin arrays are both challenging and rewarding in helping to reveal the generic magnetic features of such systems \cite{alloul09}.

Previous studies revealed that in \BCPO\ the $S = \nicefrac{1}{2}$ Cu ions form zig\-zag shaped two leg ladders along the $b$-direction of the crystal lattice \cite{koteswararao07,mentre09}.
%
%
The temperature dependence of the magnetic susceptibility $\chi(T)$ indicates that the exchange interactions favour the formation of spin-singlet dimers \cite{mentre09}. Complications are expected because of the zig\-zag geometry of the ladders \cite{koteswararao07}, which allows for frustration effects due to competing exchange interactions along the legs. Spin holes are achieved by replacing Cu with divalent $S = 0$ Zn atoms.

In previous experiments on powder samples, \BCPOd\ was studied by magnetization \cite{koteswararao07}, neutron-scattering \cite{mentre09} and nuclear-magnetic-resonance (NMR) \cite{bobroff09} measurements. Comparisons of the data with results of appropriate model calculations 
provided sets of exchange interaction parameters parallel and perpendicular to the ladder legs. Results from zero-field muon-spin relaxation ($\mu$SR) experiments were interpreted to indicate some kind of spin freezing at $x$-dependent, but low temperatures for samples with $x > 0$ \cite{bobroff09}.

In this Letter we report the results of measurements probing the static magnetization and the $^{31}$P NMR response of \textit{single crystals} 
of \BCPOd. The NMR data for $x > 0$ provide the first clear experimental evidence for the formation of local impurity-induced magnetization and its temperature dependence in such systems. Results of numerical model calculations provide support that our observations reflect the generic nature of the magnetic features of such spin arrays.

\begin{figure}[bth]
\includegraphics[width=0.40\textwidth]{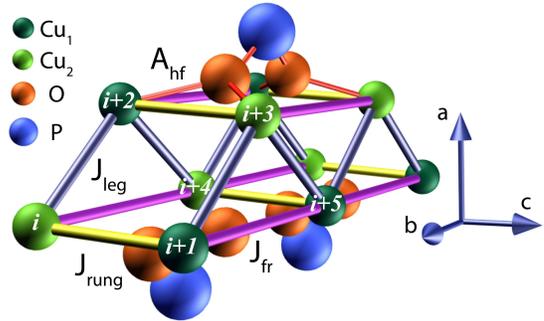} 
\caption{\label{fig:struct} (Color online) Structural subunit of \BCPO\ showing two zig\-zag Cu$^{2+}$ spin chains with frustration along the leg, coupled via rung exchange interactions. Each $^{31}$P nucleus is coupled to four Cu ions via transferred hyperfine interactions mediated by oxygen atoms, as shown for the top P atom.}
\end{figure}

High-quality single crystals of \BCPOd\ with $x$ between 0 and 0.05 were prepared as will be described in Ref.~\onlinecite{wang10}. Both phase purity and structural quality were checked via Rietveld refinement of x-ray diffraction data. For $x = 0$, the low density of defects on the Cu sublattice was confirmed by very small values of the magnetic susceptibility $\chi$ at low temperatures. For $T < 20$~K Curie-type contributions to $\chi(T)$, proportional to $x$, reflect the partial replacement of Cu by Zn. In Fig.~\ref{fig:struct} we show schematically 
parts of the structural unit cell which emphasizes the zig\-zag array of Cu sites, the essential environment around the P sites and the different couplings that were considered in our calculations.

A first magnetic characterization of the specimens invoked measurements of the magnetic susceptibility $\chi(T)$ using a SQUID magnetometer between 4 and 300 K in external fields between 0.5 and 5 T. The NMR experiments, covering temperatures between 2.6 and 300 K, probed the ensemble of $^{31}$P nuclear spins in an external field of 7.069 Tesla, corresponding to a resonance frequency of 121.828 MHz. 
The $I = \nicefrac{1}{2}$ $^{31}$P nuclei couple to the Cu atoms via transferred hyperfine interaction $A_{\mathrm{hf}}$ and hence serve as probes of the local magnetization and its dynamics.
Samples were aligned with their $b$ axis parallel to the applied field direction.
The spectra and the spin-lattice relaxation times $T_1$ were obtained with standard pulse techniques. The typical duration of the $\pi/2$ rf pulses was 5 $\mu$s. At all temperatures the $^{31}$P resonance signals were reconstructed by sweeping the frequency, using echo detection and subsequent Fourier transformation. 
In the case of $x=0$, single-pulse detection was used.
The relaxation times $T_1$ were obtained by analyzing the recovery of the magnetization after its destruction by an aperiodic comb of rf pulses.

The $\chi(T)$ data of our $x = 0$ sample are independent of the external magnetic field, at least up to 5 T. They confirm in part previously published results \cite{koteswararao10}, but with a much smaller residual $\chi$ value at low temperatures ($\sim 10^{-4}$ cm$^3$/mole Cu), reflecting the high perfection of the Cu sublattice in the parent-compound specimen. 

\begin{figure}[t]
\includegraphics[width=0.45\textwidth]{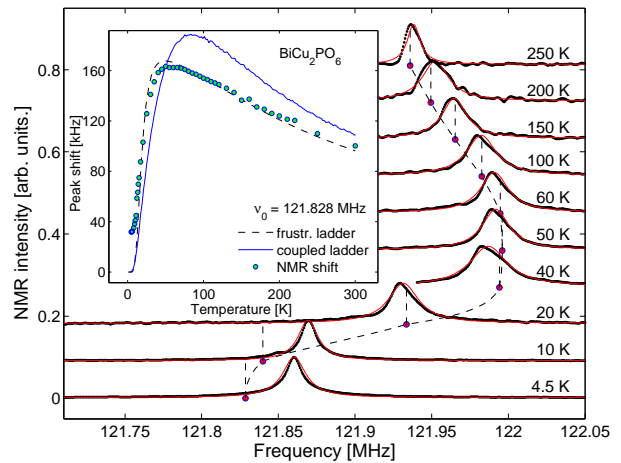} 
\caption{\label{fig:peaks} (Color online) Temperature dependence of the $^{31}$P NMR spectra of \BCPO. Red lines are Lorentzian fits to the signals. The inset shows the tem\-per\-a\-ture-in\-duced shift of the frequency at resonance maximum (open circles). The dashed lines, connecting the dots in the main frame and in the inset, represent the result of exact diagonalization on a $10\times2$ system, while the solid line results from a QMC calculation, as outlined in the text.}
\end{figure}

A selection of NMR signals recorded at different temperatures is shown in Fig.~\ref{fig:peaks}. 
The line widths (FWHM) of the individual resonances, to a fair approximation of Lorentzian shape, are of the order of 10 kHz at all temperatures. This confirms the very low concentration of remnant paramagnetic centers, i.e., the close to complete singlet formation of the Cu spins. The well preserved periodicity of the spin arrays encourages the application of exact diagonalization (ED) methods in model calculations to be discussed below. The inset of Fig.~\ref{fig:peaks} displays the temperature dependence of the line shift $K$ between 4 and 300 K, reflecting the spin susceptibility $\chi_{\mathrm{s}}(T)$. From the comparison of $K(T)$ with the $\chi(T)$ data we obtain $A_{\mathrm{hf}} = 0.50(3)$ T/$\mu_{\mathrm{B}}$. 

All our calculations to be described below are based on the Hamiltonian (see Fig.~\ref{fig:struct}):
%
\begin{align}
\label{eq:hamilt}
\mathscr{H} = \sum_{i} & \left( J_{\mathrm{leg}} \, \bm{S}_{i}\cdot \bm{S}_{i+2} + 
J_{\mathrm{rung}} \, \bm{S}_{2i-1}\cdot \bm{S}_{2i} \right. \nonumber \\
& + \left. J_{\mathrm{fr}} \, \bm{S}_{i}\cdot \bm{S}_{i+4} + g \mu_{\mathrm{B}} \,\bm{S}_{i}\cdot {\bm{H}}\right). 
\end{align}
\begin{figure*}
\includegraphics[width=0.95\textwidth]{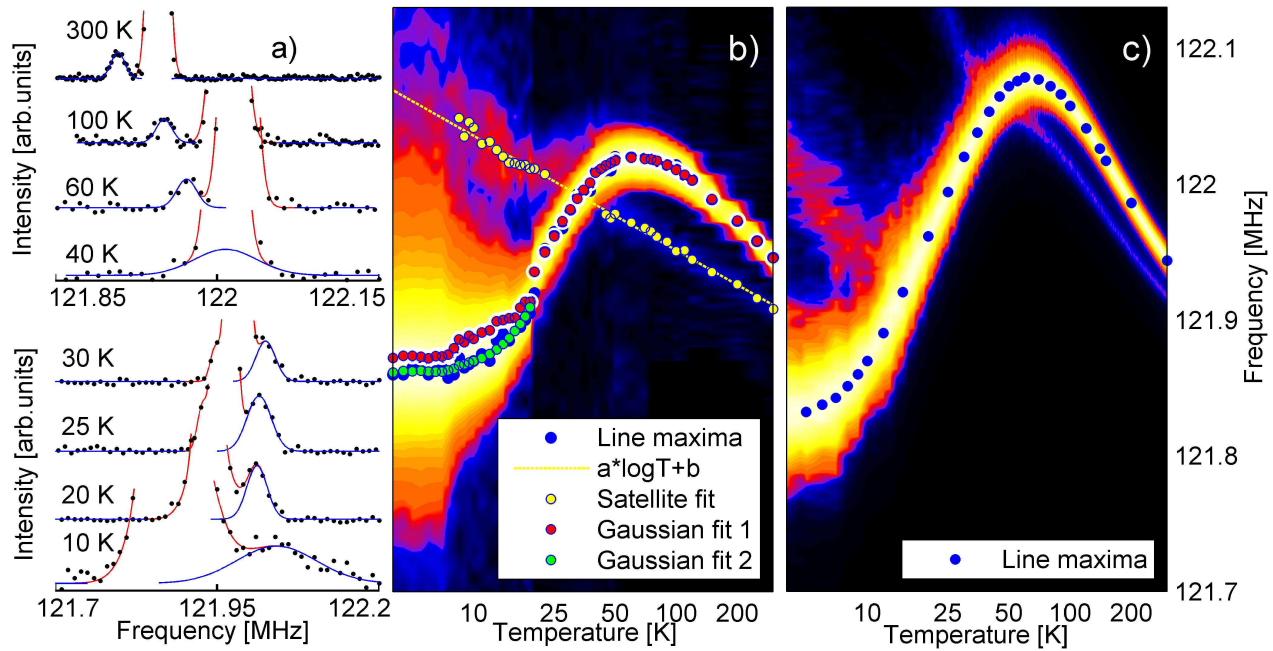} 
\caption{\label{fig:peaks_d1} (Color online) 
(a) $^{31}$P NMR spectra $I(\nu)$ for Bi(Cu$_{0.99}$Zn$_{0.01}$)$_2$PO$_6$ at different temperatures. Note the shift of the satellite signal from the low to the high frequency side of the main signal upon reducing the temperature and the substantial broadening of the main signal below 20 K. (b) Temperature dependence of the $I(\nu)$ maxima and widths; the symbols are explained in the figure. In order to guarantee the visibility of the satellite signals, the main peaks were scaled to 1 (which gives rise to an abrupt change in color intensity at $\sim 20$~K). (c) The result of a QMC simulation (for details see text).}
\end{figure*}
With respect to the data shown in Fig.~\ref{fig:peaks} we calculated the magnetization of a $10\times2$ spin system with the mentioned ED procedure, employing periodic boundary conditions and considering NNN interactions.
Due to the availability of experimental parameters, the numerical result depends merely on the choice of the exchange-parameter values. Employing the same parameters for antiferromagnetic coupling as reported in earlier work based on calculations on a $6\times2$ spin system \cite{mentre09}, namely $J_{\mathrm{leg}}/k_{\mathrm{B}} = 137.8$ K, $J_{\mathrm{rung}}/k_{\mathrm{B}} = 73.3$ K and $J_{\mathrm{fr}}/k_{\mathrm{B}} = 58.4$ K, we obtain the curve shown as dashed lines both in the main frame and the inset of Fig.~\ref{fig:peaks}. The agreement with the experimental data is quite good at high temperatures, although not perfect overall. Indeed, the same calculations on systems comprising between $6\times2$ and $10\times2$ spins revealed considerable size effects \cite{casola10}. With 12 spins the singlet-triplet gap (at 7 T) was found to be overestimated by $\sim 23$\% with respect to the thermodynamic limit (4.54 meV extrapolated value). However, with 20 spins the error reduces rapidly to 3.4\%.
Another type of calculation, using a Hamiltonian for \textit{unfrustrated} but coupled ladders, which was promoted in Ref. \cite{bobroff09}, results in the solid line also shown in the inset of Fig.~\ref{fig:peaks}. The parameters here are $J_{\mathrm{leg}}/k_{\mathrm{B}}  = J_{\mathrm{rung}}/k_{\mathrm{B}}  = 100$ K, $J_{\mathrm{interladder}}/k_{\mathrm{B}}  = 10$ K \cite{bobroff09}, and the $A_{\mathrm{hf}}$ value quoted above. It is obvious that experiment and calculation do not match in the latter case.

Next we consider the results of our investigation of samples with $x > 0$. As reported before \cite{bobroff09}, Zn impurities on Cu sites affect the resonance signals in the form of considerable broadening at low temperatures. We note, however, additional peaks (or satellites) in the spectrum, clearly visible in Fig.~\ref{fig:peaks_d1}a, where a selection of resonances for $x = 0.01$ at different temperatures is shown. The most intriguing feature is the relative shift of the satellite across the main resonance upon varying the temperature. For low concentrations $x$, the intensity ratio is $I_{\mathrm{sat}}/I_{\mathrm{tot}}=C_1^n x (1-x)^{n-1} \approx nx$, with $n$ the number of affected P nuclei around a Zn impurity. This ratio is 0.04 and 0.19 for $x = 0.01$ and 0.05, respectively. Therefore we conclude that each Zn impurity changes the average local magnetization at four neighboring P sites. The dominance of the coupling of each individual P nucleus to 4 Cu spins with $S = \nicefrac{1}{2}$ is consistent with the proximity map for P in the crystal structure of \BCPO\ and agrees with a claim in previous work based on symmetry arguments \cite{alexander10}. Model calculations to be discussed below confirm that the satellite signal is indeed due to impurity-induced changes of the local magnetization. 
We note that the satellite signal is clearly observed also in the $x = 0.05$ case (not shown). 
The availability of single-crystal samples proved crucial for the direct detection of a well resolved impurity signal, in contrast with the broad and mostly featureless resonances found in previous NMR studies on powder samples \cite{bobroff09,alexander10}.

Figure~\ref{fig:peaks_d1}b summarizes the temperature dependencies of the NMR spectrum $I\left( \nu \right)$ for $x = 0.01$ with respect to position and width. One and two Gaussians were needed to fit the main signal above and below 40~K, respectively. The onset of a second Gaussian below 40~K is most likely due to enhanced correlations in parts of the spin system. The satellite signal is again of Gaussian shape with, like the main line, a substantial broadening at low temperatures. Quite intriguing is the temperature dependence of the satellite position on the frequency axis which can be cast into the form $\nu\,[\mathrm{MHz}] = a\log T\,[\mathrm{K}]+ b$, with $a =-0.0365$, and $b = 122.12$. 

The generic nature of these observations is indicated by the results of a simulation of a $2\times100$ spin array with a QMC stochastic series expansion (SSE) algorithm and 2 spin holes located at 25 arbitrary positions. Since QMC cannot account for frustration effects, the result shown in Fig.~\ref{fig:peaks_d1}c is to be considered as preliminary. Instead of employing a time-demanding finite temperature density matrix renormalization group procedure, we aimed at showing that the shift of the satellite peak across the main line is a general feature, in agreement with the onset of staggered magnetization in the neighborhood of an unpaired spin.
The exchange couplings for the \textit{unfrustrated} (i.e.\ $J_{\mathrm{fr}} = 0$) Hamiltonian \eqref{eq:hamilt} were taken from Ref.~\onlinecite{koteswararao07}, namely $J_{\mathrm{leg}}/k_{\mathrm{B}} = 80$ K,  $J_{\mathrm{rung}}/J_{\mathrm{leg}} = 0.87$. The average of $\left< {S_i}^z\right>$ over 4 NNN Cu sites provides the local $^{31}$P field. 
We note a distinct departure of the calculated satellite shift from the observed logarithmic temperature dependence at low temperatures. As preliminary density matrix renormalization group calculations indicate, this discrepancy is most likely due to an overestimate of the induced local magnetization with the chosen Hamiltonian. 

\begin{figure}[t]
\includegraphics[width=0.48\textwidth]{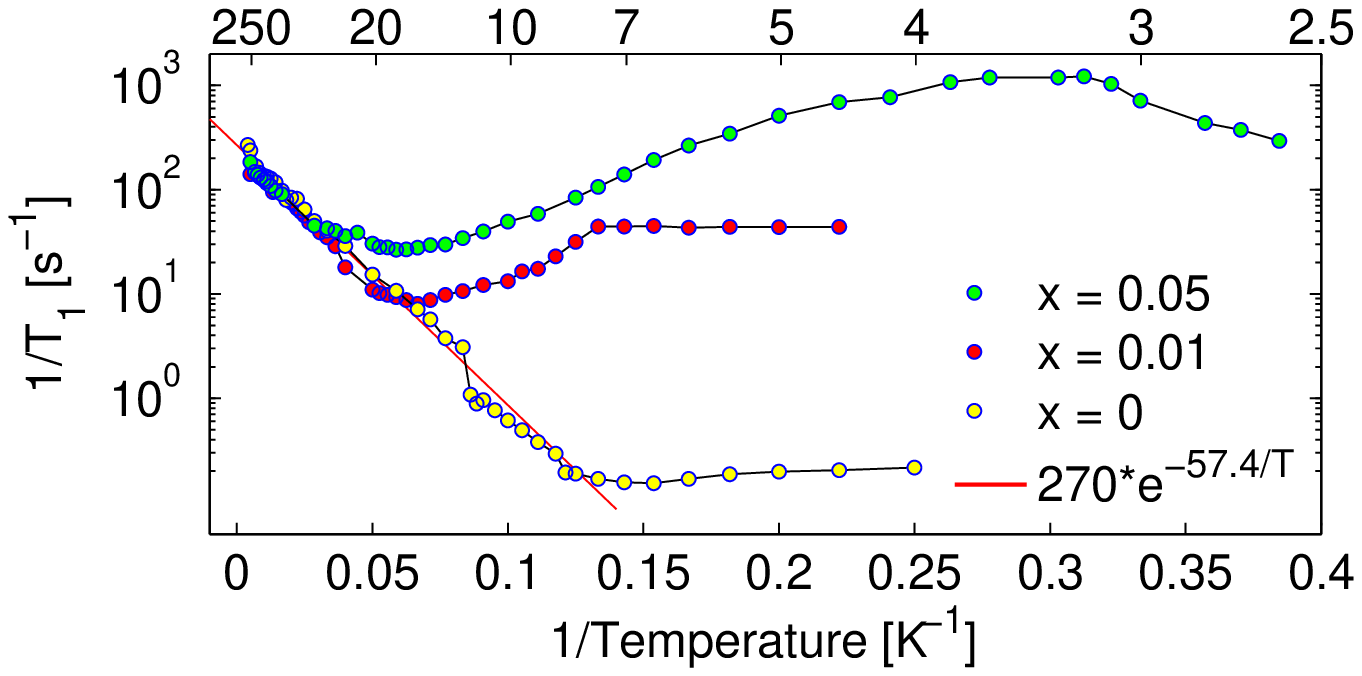} 
\caption{\label{fig:T1_relax} (Color online) $T_1^{-1}(T)$ for \BCPOd\ with $x = 0$, 0.01 and 0.05, respectively. 
}
\end{figure}

The results of our measurements of the spin-lattice relaxation rates of all three samples are shown in Fig.~\ref{fig:T1_relax}. Above 20 K, $T_1^{-1}(T)$ is practically the same for all three samples, exhibiting an activated $\exp(-\Delta/k_{\mathrm{B}}T)$ temperature dependence, with $\Delta/k_{\mathrm{B}} = 57.4 \pm 4$ K, persisting at least up to 200 K. The latter most likely indicates a negligible coupling between spin and lattice degrees of freedom in \BCPO. The gap value derived from $T_1^{-1}(T)$ and, according to Ref.~\cite{sachdev97}, scaled by a factor $\Delta_{1/T_1}/\Delta_{\chi} = 3/2$, corresponds to a spin excitation gap $\Delta_{\mathrm{sp}}/k_{\mathrm{B}} = 38.3 \pm 2.7$ K, close to values quoted in earlier works \cite{koteswararao07,mentre09,koteswararao10}.

Considerable differences among the samples develop at low temperatures, however. For $x = 0$, $T_1^{-1}(T)$ decreases further until it reaches a plateau and is practically independent of temperature below 5 K \footnote{A close inspection of $T_1^{-1}(T)$ for $x = 0$ reveals a distinct minimum at $T_{\mathrm{min}} = 6.5$ K and a subsequent weak increase below $T_{\mathrm{min}}$.}. For samples with $x > 0$, $T_1^{-1}(T)$ passes through an $x$-dependent minimum and increases again with decreasing temperature. This relaxation enhancement is quite prominent for $x = 0.05$, where it reaches a maximum at $T_0 = 3.2$ K, followed by a considerable reduction below $T_0$. 

The interpretation of the observed features $T_1^{-1}(T)$ is not straightforward. 
The data in Fig.~\ref{fig:T1_relax} indicate that the low-temperature NMR relaxation is completely dominated by the defect-induced uncompensated spin moments. Thus the maximum for $x=0.05$ 
in Fig.~\ref{fig:T1_relax} is entirely due to increasing and decreasing fluctuations in this subsystem of spins. 

From the experimental $\chi(T)$ data and the fair agreement with numerical calculations we confirm that \BCPO\ is a model compound hosting a system of $S = \nicefrac{1}{2}$ spins, coupled antiferromagnetically and periodically arranged on weakly interacting 2-leg ladders \cite{koteswararao07}. The previously mentioned frustration effects due to competing NN and NNN exchange interactions along the legs cannot be neglected, as already pointed out in earlier work \cite{mentre09}. The replacement of Cu atoms with Zn at the few at.\% level is reflected in the appearance of satellites in the $^{31}$P NMR spectra. We interpret them and their relative shifts as evidence for the presence of uncompensated dimers and locally enhanced magnetization near non-magnetic defects, respectively.
The logarithmic temperature dependence of the satellite shift is markedly different from the gap-determined shape of the curve reflecting the main-signal frequency. While this particular temperature dependence is not well understood, it implies that the gap in the excitation spectrum of the defect-induced magnetization regions is much reduced, or even zero.

The authors thank M.\ Kenzelmann and Ch.\ R\"{u}egg for essential input and support at the beginning of this project. Technical assistance was obtained from G.\ Allodi and M.\ Weller. The numerical simulations benefited from the availability of the ALPS libraries \cite{Albuquerque07}.
This work was financially supported in part by the Schweizerische Nationalfonds zur F\"{o}rderung der Wissenschaftlichen Forschung (SNF). The sample preparation at PSI was supported by the NCCR research pool MaNEP of SNF.

\end{document}